# Generator Controller Tuning Considering Stochastic Load Variation Using Analysis of Variance and Response Surface Method

Frank A. Ibarra
Department of Electrical Engineering
Universidad del Sinú
Montería, Colombia

Daniel Turizo
César Orozco-Henao
Department of Electrical and Electronics Engineering
Universidad del Norte
Barranquilla, Colombia

Javier Guerrero
School of Electrical Engineering & Computer Science
Washington State University
Bremerton, USA

***Abstract*— This article proposes a method for generator controller tuning in a power system affected by stochastic loads. The method uses the Analysis of Variance to detect the controllers with significant effect over the quality of the system response. Such quality is measured with an objective function defined as a weighted average of the Integral Absolute Error of each controller. The significant variables are then varied over a specified region in order to characterize the objective function through a regression model, which is then optimized. The method was applied to the system IEEE14 and the results were compared with benchmark parameters, showing better performance.**

***Index Terms*— ANOVA, AVR, controller, governor, power system, stochastic load, tuning.**

## I. Introduction

THE electric power networks are systems capable of transporting electric energy from the generation point to the consumption point. The network can transmit and distribute the energy but it is unable to store it for future use, for that reason a balance between generated power and demanded power must exist at every time. Power balance can be achieved by implementing automatic controllers that continuously modify the output of the generators. These controllers have parameters that must be tuned in order to ensure a proper transient response of the generators. In this article, the Automatic Voltage Regulators (AVR) and the speed governors are considered [1].

Various methods for tuning the generator controllers have been studied in the literature. If the generator-network interaction is neglected, classic analytical methods like Ziegler-Nichols can be applied [2]. Other analytical methods focused specifically on the tuning of a generator controller have been developed too [3], [4].

Analytical methods are computationally inexpensive, but the assumption of no generator-network interaction is not realistic. Fortunately, computational capabilities have grown exponentially in recent years, and research in controller tuning has focused on fine tuning using Artificial Intelligence techniques. These techniques have been used widely in different electrical engineering applications, showing improvements and advantages over classical methods [5]-[10].

An important assumption of those works is that load behaves in a deterministic way. However, real loads behave in a stochastic way and cannot be suitably modeled as being deterministic. The effect of stochastic loads in power systems has been studied in [11]-[13]. Two classic methods for the analysis and optimization of stochastic processes are the Analysis of Variance (ANOVA) and the Response Surface method [14]. Both of these methods have been successfully used in different electrical engineering applications [15]-[18].

This article proposes a methodology for generator controller tuning, tested in the benchmark IEEE14 system. In the proposed methodology, the system demands are represented as stochastic variables, in order to take into account the intrinsic variation of the load as a part of the model. The quality of the response is measured using an error function which must be minimized. First, an ANOVA is applied in order to determine which of the controller parameters have a significant effect on the mean of the objective function. Then, a factorial experiment is performed such that the samples are used to construct a regression model for the mean of the error function. This model is minimized to find the optimal parameters of the controllers (Response Surface method).

## II. Modeling

The mathematical model of a power system for transient simulations consists on a set of differential and algebraic equations (DAEs). The differential equations describe the behavior of generators and controllers, and the algebraic equations describe the behavior of the electric network.

### *A. Differential Equations*

The implemented generator model was the fourth order Two-Axis model [19], with saliency neglected. The equations of the model for generator $i$ are:

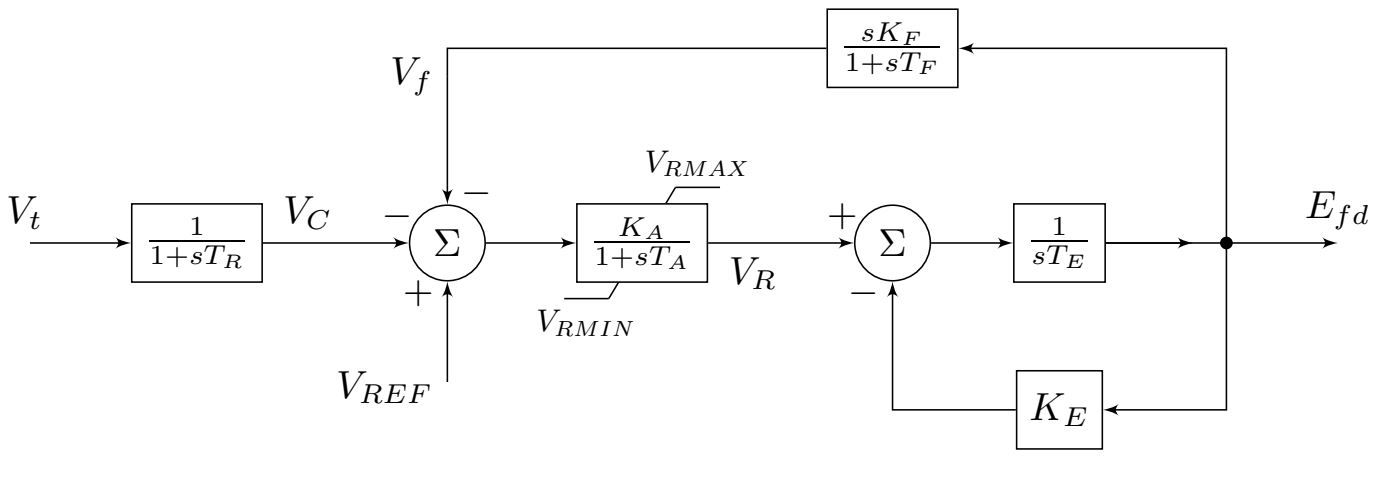

Fig. 1. AVR block diagram.

$$\dot{\theta}_i = \omega_0(\omega_i - 1) \quad (1)$$

$$\dot{\omega}_i = \frac{1}{2H_i}\Big(P_{mi} - \big(E'_{di}I_{di} + E'_{qi}I_{qi}\big) - D_i(\omega_i - 1)\Big) \quad (2)$$

$$\dot{E}'_{di} = \frac{1}{T'_{q0i}}\big(-E'_{di} + \big(X_{qi} - X'_{qi}\big)I_{qi}\big) \quad (3)$$

$$\dot{E}'_{qi} = \frac{1}{T'_{d0i}}\big(E_{fdi} - E'_{qi} - (X_{di} - X'_{di})I_{di}\big) \quad (4)$$

The meaning of the terms of these models is explained in [20]. The implemented AVR model was the Type DC1A AVR model [21], [22]. Generator saturation was neglected and so were the saturation blocks of the AVR. The block diagram of the implemented AVR model can be seen in Fig. 1. The equations of the AVR model for generator $i$ are:

$$\dot{V}_{Ci} = \frac{1}{T_{Ri}}(V_{ti} - V_{Ci}) \quad (5)$$

$$\dot{V}_{Ri} = \frac{1}{T_{Ai}}\big(K_{Ai}\big(V_{REFi} - V_{Ci} - V_{fi}\big) - V_{Ri}\big) \times u(V_{Ri} - V_{RMINi})u(V_{MAXi} - V_{Ri}) \quad (6)$$

$$\dot{E}_{fdi} = \frac{1}{T_{Ei}}\big(V_{Ri} - K_{Ei}E_{fdi}\big) \quad (7)$$

$$\dot{V}_{fi} = \frac{1}{T_{Fi}}\big(K_{Fi}\dot{E}_{fdi} - V_{fi}\big) \quad (8)$$

The governor model used was the general purpose governor presented in [23], whose block diagram can be seen in Fig. 2. The equations of the governor model for generator $i$ are:

$$\dot{u}_{0i} = \frac{1}{T_{1i}}\left(\frac{(\omega_i - 1)}{R_i} + \frac{T_{2i}}{R_i}\dot{\omega}_i - u_{0i}\right) \quad (9)$$

$$u_{1i} = \max\{\min\{P_{m0i} - u_{0i}, P_{MAXi}\}, 0\} \quad (10)$$

$$\dot{u}_{2i} = \frac{1}{T_{3i}}(u_{1i} - u_{2i}) \quad (11)$$

$$\dot{u}_{3i} = \frac{1}{T_{4i}}(u_{2i} - u_{3i}) \quad (12)$$

$$\dot{P}_{mi} = \frac{1}{T_{5i}}(u_{3i} + F_iT_{5i}\dot{u}_{3i} - P_{mi}) \quad (13)$$

### B. Algebraic Equations

The algebraic equations of a power system are the power flow equations (system has $n$ nodes where 1 to $n_G$ are PV):

$$\begin{aligned}
0 &= P_{Li}(t) - P_{Gi}(t) - \mathcal{Re}\left\{\tilde{V}_i^*\sum_{k=1}^{n} Y_{ik}\tilde{V}_k\right\}, & 1 \le i \le n_G \\
0 &= P_{Li}(t) - \mathcal{Re}\left\{\tilde{V}_i^*\sum_{k=1}^{n} Y_{ik}\tilde{V}_k\right\}, & n_G < i \le n \\
0 &= Q_{Li}(t) - Q_{Gi}(t) + \mathcal{Im}\left\{\tilde{V}_i^*\sum_{k=1}^{n} Y_{ik}\tilde{V}_k\right\}, & 1 \le i \le n_G \\
0 &= Q_{Li}(t) + \mathcal{Im}\left\{\tilde{V}_i^*\sum_{k=1}^{n} Y_{ik}\tilde{V}_k\right\}, & n_G < i \le n
\end{aligned} \quad (14)$$

Where $P_{Li}(t) + jQ_{Li}(t)$ is the complex power demand at node $i$ and time $t$, $\tilde{V}_i$ is the voltage phasor of node $i$, $Y_{ij}$ is the $ij$-th element of the admittance matrix of the system and $P_{Gi}(t) + jQ_{Gi}(t)$ is the complex power provided by generator $i$ at time $t$. The power of generator $i$ can be calculated with the generator circuit equations in $dq0$ variables (saliency neglected, time dependence omitted):

$$P_{Gi} = \mathcal{Re}\big\{\big(E'_{di} + jE'_{qi}\big)\big(I_{di} - jI_{qi}\big)\big\} \quad (15)$$

$$Q_{Gi} = \mathcal{Im}\big\{\big(E'_{di} + jE'_{qi}\big)\big(I_{di} - jI_{qi}\big)\big\} \quad (16)$$

$$I_{di} + jI_{qi} = \frac{E'_{di} + jE'_{qi} - \tilde{V}_i e^{j\left(\frac{\pi}{2} - \theta_i\right)}}{R_{ai} + jX'_{di}} \quad (17)$$

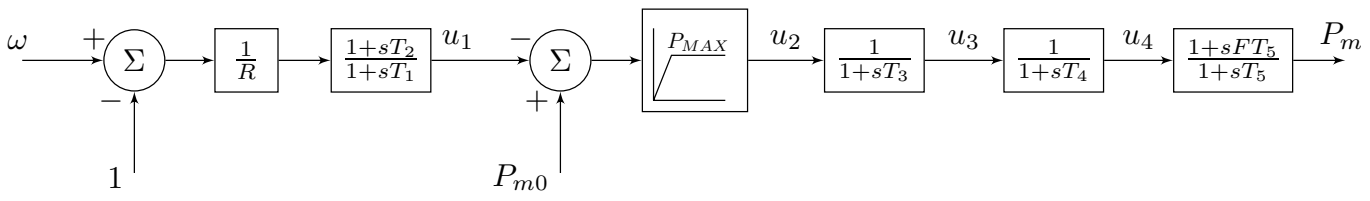

Fig. 2. Governor block diagram.

The natural variation of the system loads can be represented with the Gaussian White Noise model proposed in [11]-[13], as follows:

$$P_{Li}(t) + jQ_{Li}(t) = \big(1 + \lambda_i dW_i(t)\big)(P_{L0i} + jQ_{L0i}) \quad (18)$$

Where $P_{Li} + jQ_{Li}$ is the complex power demand at node $i$, $P_{L0i} + jQ_{L0i}$ is its mean, $\lambda_i$ is the stochastic penetration constant, and $dW_i(t)$ is a White Gaussian Noise (WGN) variable, defined as the derivative of a standard Wiener process. The previous model assumes the mean load value remains constant. The mean load may experiment increasing or decreasing trends due to its daily and weekly periodicity. In order to take into account those trends the mean load is assumed to be varying linearly:

$$P_{Li}(t) + jQ_{Li}(t) = \big(1 + \lambda_i dW_i(t)\big)(P_{L0i} + jQ_{L0i})(1 + m_i t) \quad (19)$$

Where $m_i$ is the slope of the mean. In this work, the slope is assumed to be a uniformly distributed random variable. The response of the system for a specific realization of load is calculated by numerically solving the differential equations of the system, using the trapezoidal method [1], [19].

### C. Objective Function

The quality of the system response is measured with an error function $f$. The chosen function in this article is a simplified version of the objective function presented in [10]:

$$f = \frac{\sum_{i=1}^{n_g}\Big(w_{1i}\int_0^{t_f}|V_{ti}(t) - V_{ti}(0)|dt + w_{2i}\int_0^{t_f}|\omega_i(t) - \omega_i(0)|dt\Big)}{\sum_{i=1}^{n_g}(w_{1i} + w_{2i})} \quad (20)$$

Where $n_g$ is the number of synchronous machines, $w_{1i}$ is the weight of the voltage error for machine $i$, and $w_{2i}$ is the weight of the frequency error for machine $i$. $w_{1i}$ is 1 for all machines, $w_{2i}$ is 1 for all generators, and 0 for all synchronous condensers.

TABLE I
DESIGN FACTOR LEVELS

| Design Factor | Low Level [p.u] | High Level [p.u] | Normal Operating Value | Factor Coding |
|---|---|---|---|---|
| $K_{A8}$ | 25 | 500 | 400 | A |
| $K_{A6}$ | 25 | 500 | 400 | B |
| $K_{A3}$ | 25 | 500 | 400 | C |
| $K_{A2}$ | 25 | 500 | 25 | D |
| $K_{A1}$ | 25 | 500 | 50 | E |
| $R_2$ | 0.02 | 0.1 | 0.05 | F |
| $R_1$ | 0.02 | 0.1 | 0.05 | G |

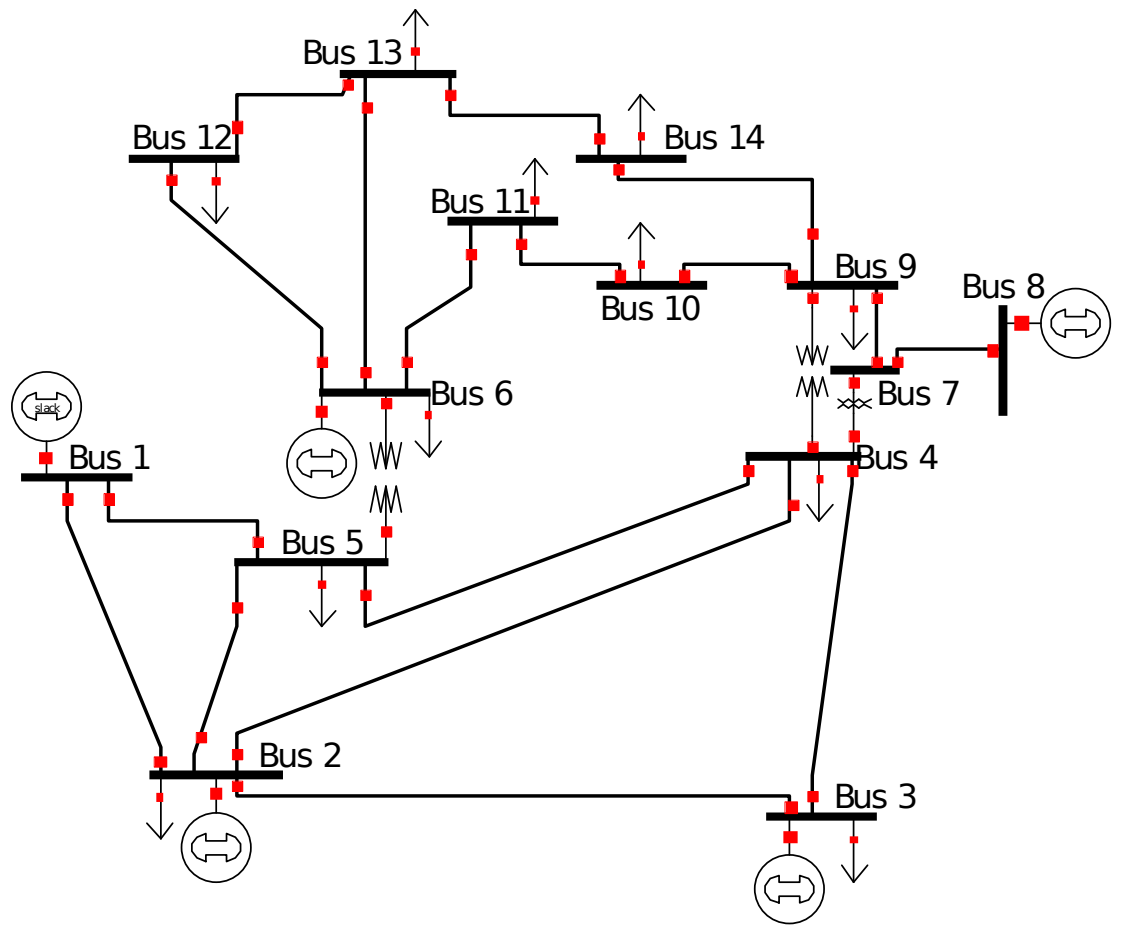

Fig. 3. Circuit diagram of system IEEE14.

## III. Experimental Design

In order to perform the optimization of the objective function, a statistical model is constructed as follows:

### *A. Screening Experiment*

A common screening experiment is the $2^k$ factorial design, in which the process is executed $2^k$ times, varying the value of the possible factors from their minimum value to their maximum value, performing all the possible combinations of factor values.

In this article, a $2^7$ factorial design with no repetitions was selected. In order to execute the experiment, the experimental region was defined by setting the levels of operation of the factors, based on the typical values they can take, which were taken from [23]. A normal probability plot of the factors and interactions is drawn to find candidate significant factors [14].

### *B. $3^k$ Factorial Design*

Using the ANOVA, the non-significant factors are fixed at a convenient value, leaving the significant factors values to be defined. The $3^k$ factorial design is an experiment similar to the $2^k$ factorial design, with the difference that an intermediate level is introduced. The values of the factors at the intermediate level are the means of the values at the low and high levels [14].

### *C. Response Surface Method*

The data collected with the $3^k$ factorial experiment can be used to construct a regression model fitting the response variable inside the experimental region. The model can be optimized with classical optimization techniques, and if the optimum point is inside the experimental region, then is also an optimum point of the response variable.

## IV. Screening Results

The system used in this article is the benchmark system IEEE14. This system is composed of 14 nodes with 5 synchronous machines, 2 of them working as generators and the rest as synchronous condensers. The original IEEE14 does not have generator or controller data, but in 2015 Demetriou *et al.* estimated the generator and controller data of this and other reference systems [21]. The controller and generator data used for the IEEE14 system can be found in [20]. The diagram of the system is shown in Fig. 3.

The tunable parameters of the controllers are the amplifier gain $K_A$ of the AVRs and the speed droop $R$ of the governors. Synchronous motors do not have any injection of mechanical power, and thus they do not possess speed governors, but they possess AVRs. The parameters $K_{Ai}$ and $R_i$ are the AVR and governor parameters of the synchronous machine connected to node $i$. Therefore, the parameters to be tuned are $K_{A8}$, $K_{A6}$, $K_{A3}$, $K_{A2}$, $K_{A1}$, $R_2$ and $R_1$.

In [20], [21] the authors define a set of operating parameters for the system IEEE14, which will be taken as the normal operating parameters. The values are in Table I.

The $2^k$ factorial experiment was performed using simulations of the system response calculated with a simulation program developed in MATLAB®. The objective function was calculated using simulations of 30 seconds, solved with the implicit trapezoidal method and a time step of 0.2 seconds. The parameters $\lambda_i$ were set at a value of 0.5% for all the loads, and the parameters $m_i$ were set to vary uniformly between ±0.2%. For simplicity, the effect of each factor is coded with a letter, (see Table I).

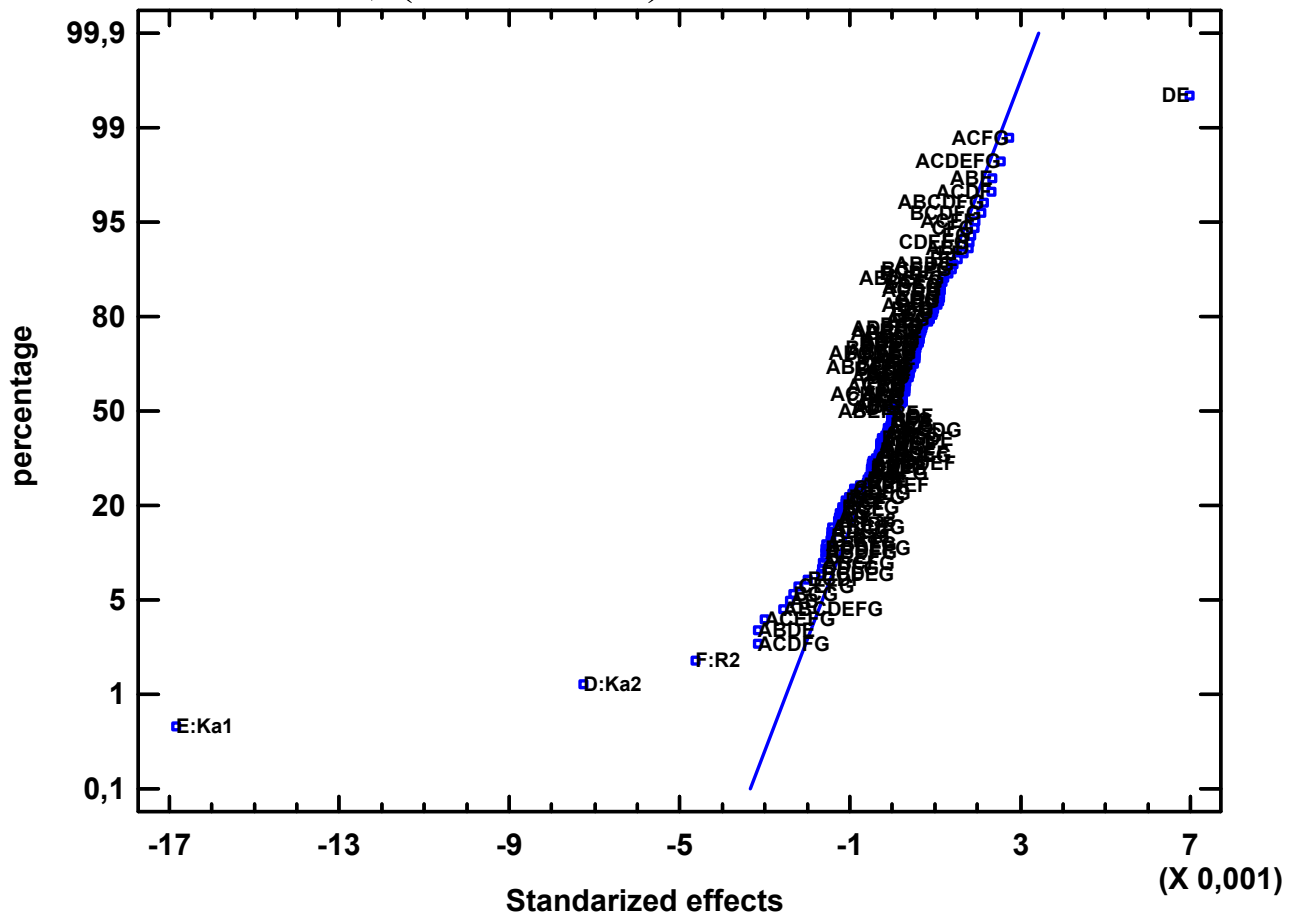

Fig. 4. Normal plot of the design effects.

From the normal plot in Fig. 4, it can be concluded that the significant effects are effects D, E, F and DE. A multifactor ANOVA was performed for these effects. In order for the ANOVA to be statistically valid, three assumptions must hold [14]:

- The variances of the levels of each factor must be equal (homoscedasticity).
- The residuals of the samples must be normally distributed.
- The residuals of the samples must be independent.

The equality of variance assumption was tested using Levene's test with significance $\alpha = 0.05$. The factor E showed significant differences in its variances, as its P-Value was 1.206E-9, lower than the significance.

In order to correct the violation of the homoscedasticity assumption, the data was transformed using the power transformation $y^* = y^\lambda$. An adequate value $\lambda = -1.3$ was found. Levene's test was applied to the significant factors using the transformed data, obtaining the results of Table II.

TABLE II
LEVENE'S TEST FOR THE SIGNIFICANT FACTORS TRANSFORMED ($\lambda = -1.3$)

| Factor | Test | P-Value |
|---|---|---|
| D | 3.5685 | 0.061184 |
| E | 3.4545 | 0.065413 |
| F | 1.1893 | 0.277553 |

As all the P-Values are greater than the significance, then it can be concluded that the homoscedasticity assumption holds. A multifactor ANOVA was then applied to the transformed data, using the same significance of Levene's test. The results are shown in Table III.

TABLE III
ANOVA TABLE FOR THE $2^7$ FACTORIAL EXPERIMENT

| SoV | SS | DoF | MS | Fo | P-Value |
|---|---|---|---|---|---|
| D | 16847.5 | 1 | 16847.5 | 21.88 | 0.0000 |
| E | 277651 | 1 | 277651 | 360.54 | 0.0000 |
| F | 20878.6 | 1 | 20878.6 | 27.11 | 0.0000 |
| DE | 5749.52 | 1 | 5749.52 | 7.47 | 0.0072 |
| Error | 94720.9 | 123 | 770.089 | | |
| Total | 415847 | 127 | | | |

It can be concluded that the effects D, E, F and DE have a significant influence over the response variable. To validate the results of the ANOVA, the other two assumptions must be verified. The residuals were calculated using the fixed effects model of the ANOVA. The Shapiro-Wilks test was performed on the residuals, obtaining a P-Value of 0.38 (normality not rejected). Independence of the residuals was also confirmed.

## V. OPTIMIZATION RESULTS

From the results of the screening experiment, the significant factors are D, E, and F ($K_{A2}$, $K_{A1}$ and $R_2$, respectively). Factors A, B, C and G ($K_{A8}$, $K_{A6}$, $K_{A3}$ and $R_1$, respectively) are not significant and therefore there is no need to change them. For this reason, the non-significant factors are set to their normal operating values. A $3^3$ factorial experiment was performed to collect the required data for the regression model. The values of the factors at the different levels are shown in Table IV.

TABLE IV
FACTOR LEVELS OF THE $3^3$ EXPERIMENT

| Factor | Low Level [p.u.] | Medium Level [p.u.] | High Level [p.u.] |
|---|---|---|---|
| $K_{A2}$ | 25 | 262.5 | 500 |
| $K_{A1}$ | 25 | 262.5 | 500 |
| $R_2$ | 0.02 | 0.06 | 0.1 |

The objective is to find a regression model able to represent the objective function over the experimental region. The proposed structure for the regression model was the following:

$$y = \beta_0 + \beta_1 K_{A2} + \beta_2 K_{A1} + \beta_3 K_{A1}^2 + \beta_4 K_{A2}^2 K_{A1} \quad (21)$$

The regression model constants $\beta_0$ to $\beta_4$ were calculated using least squares fitting. The obtained model was the following:

$$\begin{aligned} y = {} & 0.0518 - 4.5868 \cdot 10^{-5} K_{A2} \\ & -1.4479 \cdot 10^{-4} K_{A1} + 1.6678 \cdot 10^{-7} K_{A1}^2 \\ & +1.8613 \cdot 10^{-10} K_{A2}^2 K_{A1} \end{aligned} \quad (22)$$

The statistical significance of the model was verified by applying an ANOVA. The results are shown in Table V. The statistical significance of each term of the model was verified too. The results are shown in Table VI.

TABLE V
ANOVA FOR THE SIGNIFICANCE OF THE REGRESSION MODEL

| SoV | SS | DoF | MS | Fo | P-Value |
|---|---|---|---|---|---|
| Model | 0.00280 | 4 | 0.00070 | 7.5615 | 0.0005 |
| Error | 0.00203 | 22 | 9.26E-5 | | |
| Total | 0.00483 | 26 | | | |

TABLE VI
ANOVA FOR THE SIGNIFICANCE OF THE MODEL TERMS

| Term | Coeficient | LB 95% | UB 95% | to | P-Value |
|---|---|---|---|---|---|
| Constant | 0.051822 | 0.040386 | 0.063258 | 9.39 | 3.7E-9 |
| $K_{A2}$ | -4.58E-5 | -7.74E-5 | -1.43E-5 | -3.01 | 0.0064 |
| $K_{A1}$ | -0.000144 | -0.000226 | -6.41E-5 | -3.72 | 0.0012 |
| $K_{A1}^2$ | 1.66E-7 | 2.23E-8 | 3.11E-7 | 2.39 | 0.0256 |
| $K_{A2}^2 K_{A1}$ | 1.86E-10 | 8.02E-12 | 3.64E-10 | 2.16 | 0.0413 |

The model does not include any term involving $R_2$, which may seem strange because the screening experiment labeled it as significant. The reason is that various other models including the term $R_2$ were tested, and all of them showed that the term is not significant. Furthermore, the selected model presented here has been proven to be statistically valid. Finally, as the model is only valid inside the experimental region, the optimization can be formulated as:

$$\min_{[K_{A2}\ K_{A1}]} y \quad (23)$$

subject to:

$$25 \leq K_{A2} \leq 500 \quad (24)$$

$$25 \leq K_{A1} \leq 500 \quad (25)$$

The optimization problem was solved, obtaining the optimal set of parameters shown in Table VII (note that the others factors were kept at their normal values).

TABLE VII
OPTIMAL SET OF CONTROLLER PARAMETERS

| Parameter | Optimal Value [p.u.] |
|---|---|
| $K_{A8}$ | 400 |
| $K_{A6}$ | 400 |
| $K_{A3}$ | 400 |
| $K_{A2}$ | 330.10 |
| $K_{A1}$ | 373.26 |
| $R_2$ | 0.05 |
| $R_1$ | 0.05 |

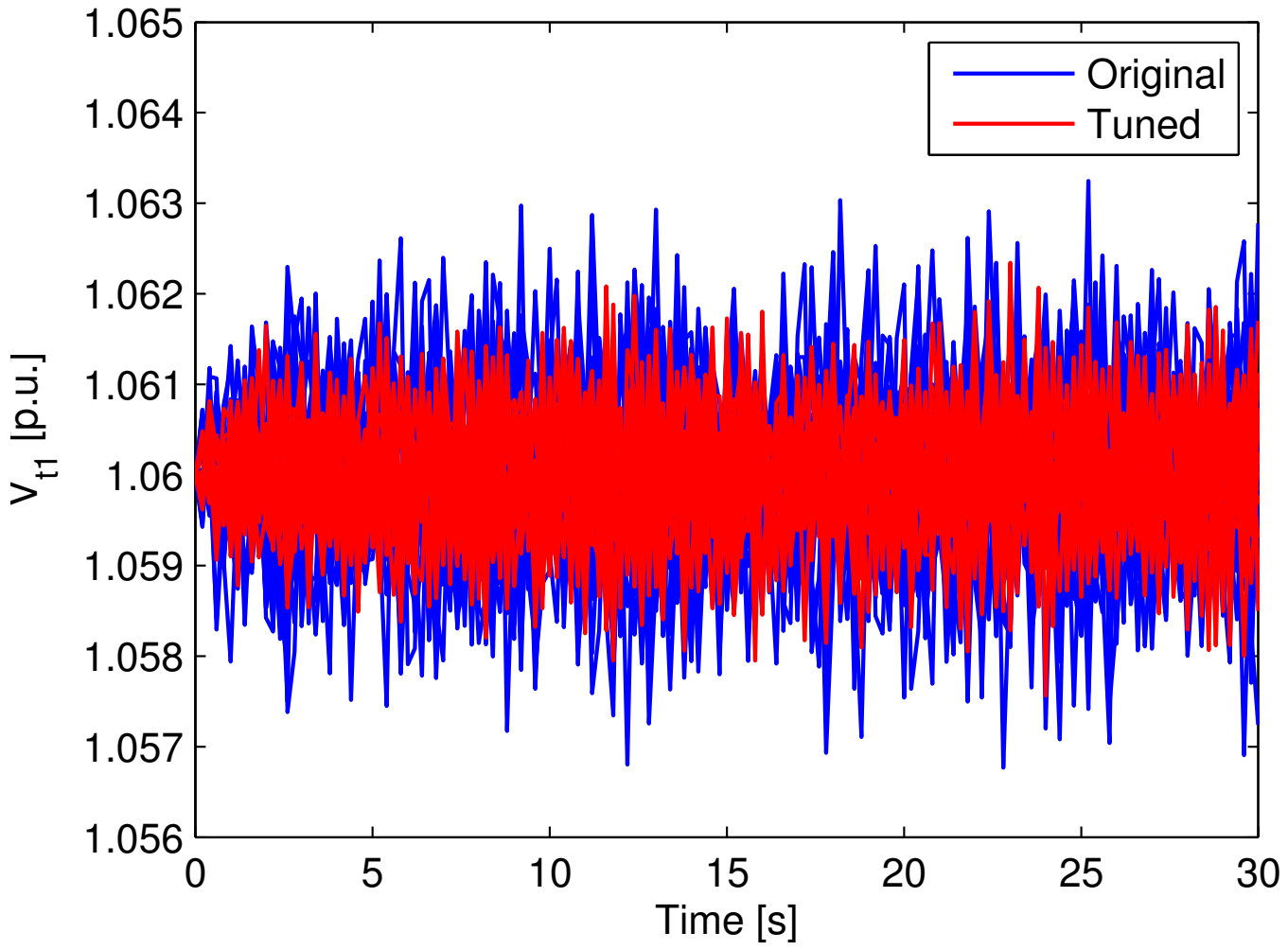

Fig. 5. Observations of Slack node voltage response (20 samples).

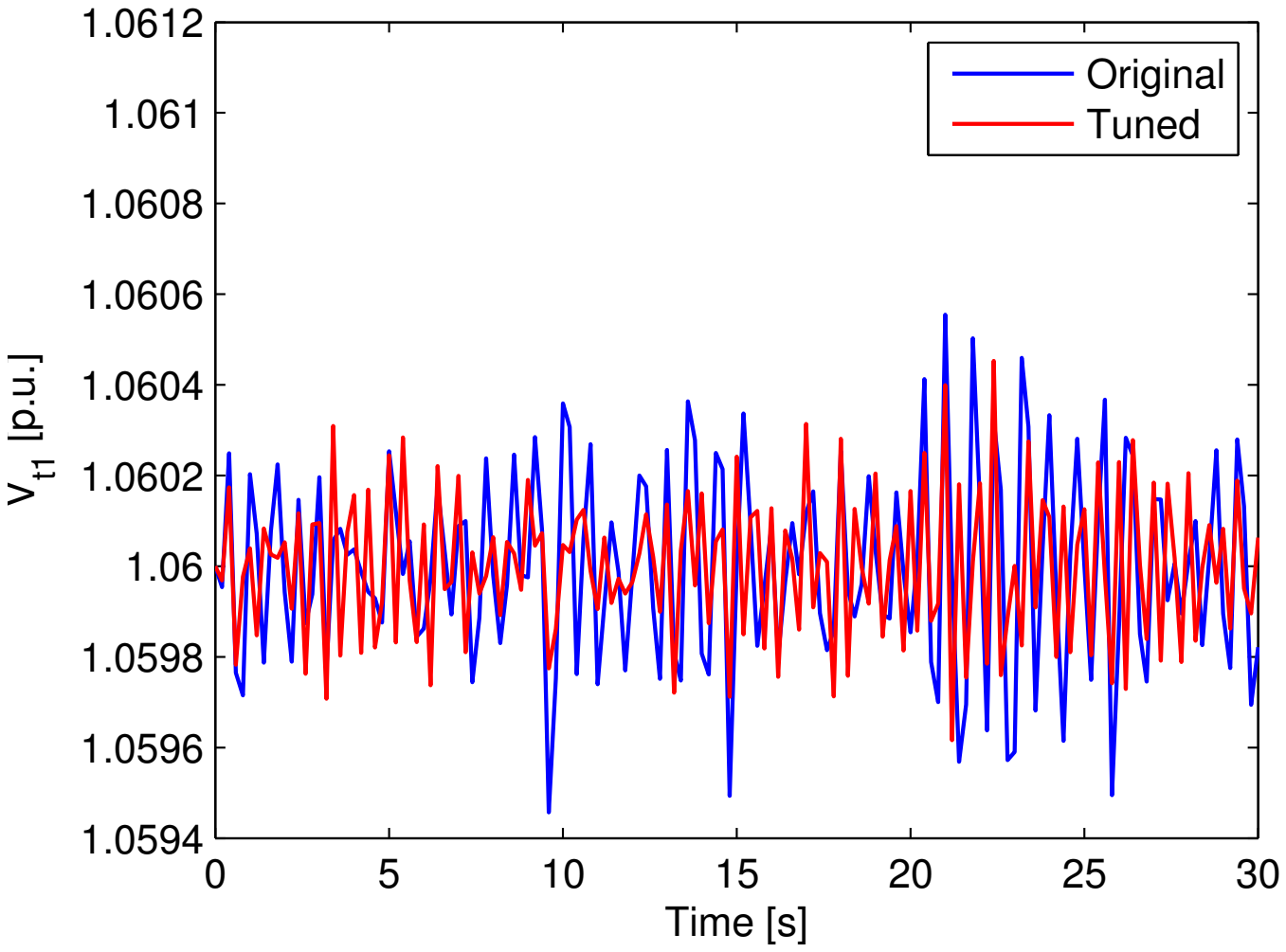

Fig. 6. Average Slack node voltage response (20 samples).

Fig. 5 shows a set of 20 observations of the voltage response of the slack node with both the original and the tuned set of parameters. Fig. 6 shows the average response for both sets of parameters. From the figures it can be seen that there is a significant reduction of the voltage variations, but it is necessary to validate the generality of these conclusions using statistical methods.

## VI. Validation

In order to validate the optimal set of parameters, they were compared against the normal parameters of [20]. To do this, a hypothesis test for the difference of means was performed. 20 samples of the objective function with each set of parameters were taken. A hypothesis test for the difference of means was performed (homoscedasticity was confirmed), with the results shown below:

TABLE VIII
Hypothesis Test for the Difference of Means

| Hypothesis test of means | | | |
|---|---|---|---|
| to | P-Value | α | Conclusion |
| 7.2545 | 5.58E-9 | 0.05 | $H_0$ Rejected |

The null hypothesis is rejected, as shown in Table X. Thus the mean of the objective function with the optimal set of parameters is lower than the mean with the operating set of parameters.

## VII. Conclusions

This article describes a method for the tuning of generator controllers in a power system affected by stochastic loads, by applying the Analysis of Variance and the Response Surface Method. The method was applied to the system IEEE14. The regression model obtained from the experiments was minimized, and the optimal set of parameters obtained was compared against the typical set of parameters, showing better performance.